\documentclass{aa}  
\usepackage{graphicx}
\usepackage{color}
\usepackage{txfonts}
\usepackage{natbib}

\begin{document} 

   \title{Reconstructing solar magnetic fields from historical observations}

   \subtitle{V. Sunspot magnetic field measurements at Mount Wilson Observatory}

\author{Alexei A. Pevtsov\inst{1,}\inst{2}
          \and
          Kseniya A. Tlatova\inst{3}
          \and
          Alexander A. Pevtsov\inst{1}
          \and
                Elina Heikkinen\inst{2}
        \and
        Ilpo Virtanen\inst{2}
        \and
                Nina V. Karachik\inst{4}
        \and
        Luca Bertello\inst{1}
        \and
        Andrey G. Tlatov\inst{5,}\inst{6}
        \and
        Roger Ulrich\inst{7}
        \and
        Kalevi Mursula\inst{2}
          }

\institute{National Solar Observatory, Boulder, CO 80303, USA\\
             \email{apevtsov@nso.edu}; \email{lbertello@nso.edu}; \email{aapevtsov@nso.edu}
\and
ReSoLVE Centre of Excellence, Astronomy and Space Physics research unit, University of Oulu, POB 3000, FIN-90014, Oulu, Finland\\
              \email{Elina.Heikkinen@oulu.fi}; \email{Ilpo.Virtanen@oulu.fi}; \email{kalevi.mursula@oulu.fi}
         \and
Central Astronomical Observatory
of the Russian Academy of Sciences at Pulkovo,
Saint Petersburg, 196140, Russia
         \email{k.tlatova@mail.ru}
         \and
         Uzbekistan Academy of Sciences, Ulugh Beg Astronomical Institute, Tashkent, Uzbekistan\\
         \email{ninakarachik@mail.ru}
         \and
         Kislovodsk Mountain Astronomical Station of Pulkovo Observatory, Kislovodsk, Russia\\
         \email{tlatov@mail.ru}
         \and
         Kalmyk State University, Elista, Russia
         \and
                University of California at Los Angeles (UCLA), Los Angeles, CA, USA\\
        \email{ulrich@astro.ucla.edu}
             }

   \date{Received ; accepted}

 
  \abstract
  {Systematic observations of magnetic field strength and polarity in 
sunspots began at Mount Wilson Observatory (MWO), USA in early 1917. Except for a 
few brief interruptions, this historical dataset has continued until the present.}
   {Sunspot field strength and polarity observations are critical in our project of reconstructing the solar magnetic field over the last hundred years. We provide a detailed description of the newly digitized dataset of drawings of sunspot magnetic field observations.}
   {The digitization of MWO drawings is based on a software 
   package that we developed. It 
   includes a semiautomatic selection of solar limbs and other 
   features of 
   the drawing, and a manual entry of the time of observations, measured field strength, and other notes  handwritten on 
   each drawing. The data are preserved in an  MySQL database.}
   {We provide a brief history of the project and describe the results from 
digitizing this historical dataset. We also provide a summary of the 
final dataset and describe its known limitations. Finally, we compare the sunspot magnetic field 
measurements with those from other instruments, and demonstrate that, if needed, the 
dataset could be continued using modern observations such as, for example, the 
Vector Stokes Magnetograph on the Synoptic Optical Long-term 
Investigations of the Sun platform.}
   {}
   {}
   \keywords{Sun: magnetic fields -- sunspots -- History and philosophy of astronomy -- Astronomical data bases}
\titlerunning{Reconstructing solar magnetic fields from historical observations. V}
\authorrunning{A.A. Pevtsov et al.}
\maketitle
%

\section{Introduction}\label{sec:intro}

In 1896, the Dutch physicist Pieter Zeeman conducted a series of experiments
that led to the discovery that in the presence of strong magnetic fields
some spectral lines split into multiple components. Later the same year,
Hendrik Lorentz proposed a theoretical explanation of these observations 
based on his theory of electromagnetic radiation. In 1902, Zeeman and 
Lorentz shared the Nobel Prize for the discovery and explanation of a 
fundamental physical effect we now refer to as the Zeeman effect. In 
the concluding section of the article describing his experiments 
on spectral line splitting and polarization, Zeeman posed a
question about a possible effect of then hypothesized magnetic 
fields on the Sun  
\citep{Zeeman1897}. In 1908, American astronomer George E. 
Hale \citep{Hockey.etal2014} discovered the presence of the 
magnetic field in sunspots.
Figure \ref{fig:zef}a shows an example of a spectral line 
splitting in the presence of a strong magnetic field in 
a sunspot.
While the discovery of the magnetic field in sunspots is 
usually credited to Hale's 1908 paper \citep{Hale1908}, it 
was, in fact, a combination of
formal articles and personal communications (including short telegrams) 
between Hale and Zeeman
that led to a discovery of both the longitudinal (along  the
line of sight) and transverse (orthogonal to the line of sight) 
magnetic fields in sunspots. The 
broadening pattern in spectral lines characteristic to the Zeeman 
effect was first observed by other astronomers \citep[e.g.,][]{Mitchell1906}, 
who, unlike Hale, failed to properly interpret the observations in 
terms of magnetic fields. Additional information on the history of 
the discovery of magnetic fields on the Sun is provided in
\citet{Bakker1946,Spencer1965,Harvey1999,Iniesta1996,Stenflo2015}.

Early observations conducted in 1908  
indicated that a proper measurement of sunspot magnetic fields
requires a telescope with a large image size in the focal plane 
(which implies a longer focal length), and a spectrograph with a high resolving power. In 1909, the Carnegie Institution of Washington 
provided the necessary funding for the construction of a new 150 feet (45.7 meters) tower telescope at the Mount Wilson Observatory (MWO) 
in California. The construction of the tower was completed in 1910 and the first 
observations were taken in 1911. The construction project was finished in 
May 1912 with the completion of the Littrow spectrograph 
\citep{Howard1985}. Meanwhile, studies of the magnetic
field on the Sun continued with such important findings as a possible height dependence of 
sunspot field strengths, the recognition of the bipolar structure of active 
regions \citep{Hale1910}, and even with a claim about the presence of a global dipolar magnetic field on the Sun \citep{Hale1913}. The 
validity of the latter finding was questioned by some later researchers 
\citep[e.g.,][]{Stenflo2015} and, indeed, based on current knowledge, 
the mean strength of the polar field derived by Hale may appear too large. 
Nevertheless, the measurements showed a 
hemispheric asymmetry in the polarity of 
magnetic fields at middle and high latitudes. 
Since these field values were longitudinal averages over a range 
of latitudes, we could speculate that these measurements might represent the 
imbalance of large-scale 
magnetic fields of decaying active regions. 
Then the derived field strengths 
may not be so unreasonable after all.

With the completion of the 150 feet solar tower telescope, the MWO synoptic
program of daily sunspot drawings including the 
measurements of sunspot 
polarities and field strengths began in 1917 with the first measurements made by 
Ferdinand Ellerman \citep{Hockey.etal2014} on 4 January  1917. With a few 
interruptions 
due to funding issues, this program has continued until the present. In the 
fall of 1978, water from severe rainfalls flooded the spectrograph and 
partially damaged the spectrograph grating and the lens. As a 
precaution, all following gratings were installed in a 
watertight box. In 1984, the Carnegie Institution of Washington made a 
decision to close the MWO. In late 1985, a 
memorandum of understanding was signed between the Carnegie 
Institution and the University of California in Los Angeles (UCLA), 
allowing the continuation of the sunspot drawing program. 
The primary objective of the funded program was the study of dynamical 
structures such as the torsional oscillations, which required multiple 
magnetogram or Dopplergram observations throughout the day. During the peak of 
the solar cycle, the drawings occupied two hours or more to complete,  leading 
to the loss of observations for the primary program.  Consequently, daily 
observations were stopped on 15 September 2004. In 2005, only nine drawings were 
made, and in 2006, there were only two drawings. The daily observations 
restarted on 25 January 2007 with the compromise that the magnetic field in 
the smaller spots would not be measured.
The most recent observations, however, are made 
on a volunteer basis by longtime observer Steve Padilla. 
Thus, they could stop at any time.

Manual measurements of sunspot field strengths using the approach 
pioneered by Hale were conducted at several other observatories, such as Potsdam, Rome, and the Crimean Astrophysical Observatory (CrAO) 
\citep[see][]
{Livingston.etal2006}. Another notable synoptic dataset comes from the
network of observatories in the former Soviet Union, which 
operated within the framework of 
the Solar Service program \citep{Pevtsov.etal2011}. This latter 
dataset extends from early 1950s through late 1990s. After 1998, it 
has continued based on daily observations from a single station (CrAO).

In the mid-2000s to early 2010s, a series of papers related to the long-term 
variation of sunspot field strengths rejuvenated the interest in 
the MWO synoptic dataset 
\citep{Penn.Livingston2006,Penn.Livingston2011,Pevtsov.etal2011,Watson.etal2011,Rezaei.etal2012,Nagovitsyn.etal2012,Livingston.etal2012,Pevtsov.etal2014,Lockwood.etal2014,Rezaei.etal2015,Tlatova.etal2015}. 
This renewed interest led to our present collaborative effort in 
digitizing the Mount Wilson sunspot drawings. As this project enters 
its final stage, we feel obligated to provide a detailed description 
of the resulting dataset. In Section \ref{sec:method} we describe the 
concept of measuring the magnetic fields employed by Hale and provide a 
history of instrument changes. In Section 
\ref{sec:digit} we summarize the steps taken in the course of drawing 
digitization and outline the features of the  
searchable database of sunspot field strengths that we created. Sections 
\ref{sec:issues}--\ref{sec:errors} provide critical discussions of 
data quality, the estimate of uncertainties, and the known issues. Because of an
inevitable termination of the current sunspot drawings program at MWO, a possible continuation of this time series 
using modern Stokes polarimeter data has been developed at the 
National Solar Observatory (NSO) \citep{Hughes.etal2013}. This method relies 
on measuring the Zeeman splitting in Stokes $I$ and $V$ profiles, and it 
closely mimics the original approach by Hale. Section \ref{sec:compare} 
provides a brief summary of that approach and compares the derived 
field strengths with those from the MWO manual measurements.

\section{Method and observations}\label{sec:method}

\begin{figure*} 
\centerline{\includegraphics[width=17cm,trim=0 100 0 100,clip=]{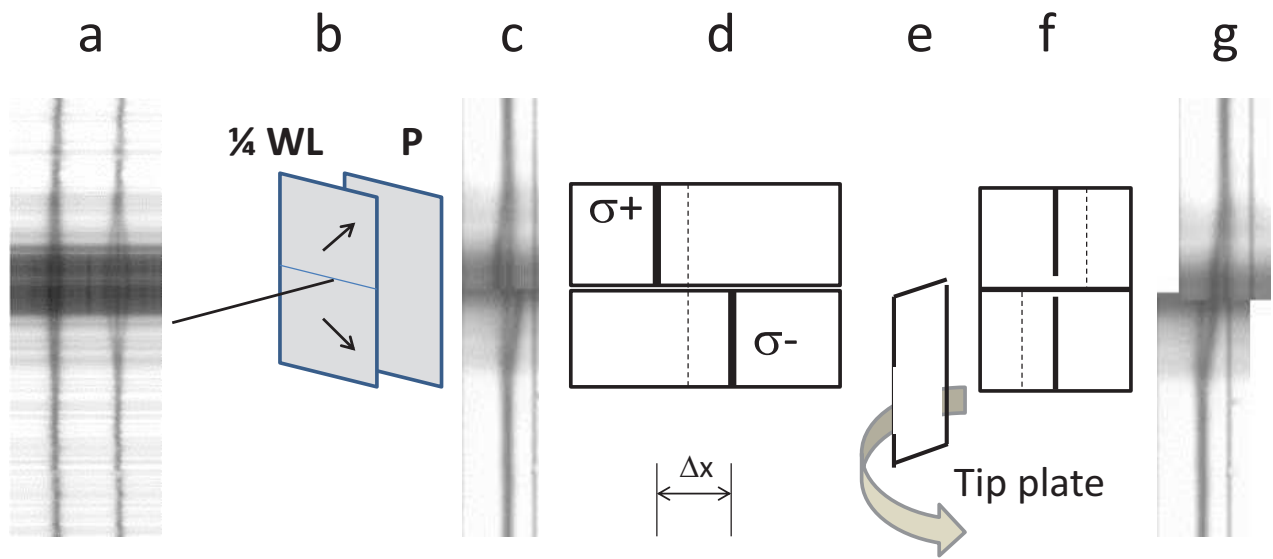}}
\caption{Principle of measuring the field strength in
sunspots. (a) Example of Zeeman splitting of spectral 
lines \ion{Fe}{I} 6301.5 \AA\ (left) and \ion{Fe}{I} 6302.5 
\AA\ (right) as observed by the spectropolarimeter on board 
the Hinode spacecraft. The horizontal direction 
corresponds to the wavelength space and the vertical 
direction shows spatial dimension (along the spectrograph slit; in this example, across a small sunspot).   
The dark diffuse horizontal band in the middle of the spectra 
corresponds to sunspot 
umbra, and two gray bands above and below this dark 
band are sunspot penumbra. Brighter areas farther 
outward from the penumbra correspond to the photosphere. 
There, the brighter and 
darker horizontal lines in panel (a) correspond to the 
granulation pattern as the spectrograph slit crosses 
(brighter) centers of granules and (darker) 
intergranular spaces.
Owing to the atmospheric seeing, the granulation pattern 
is less
visible in spectra taken with VSM/SOLIS (panels c and g).
To measure the magnetic field, two Zeeman components with left- and 
right-hand polarization are spatially separated with 
the two strips of optical retarder (quarter-
wavelength plate, 1/4 WL) and a polarizer (P). Upper 
and lower strips of retarder are oriented orthogonal 
to each other and at 45 degrees to the polarizer, as 
shown in panel (b). Each strip (retarder and polarizer) 
transmits only one of two Zeeman components, which allows  
the left-hand and right-hand 
polarized components been separated in the vertical 
direction (panel c). Panel d is a simplified, 
schematic representation of panel c (central 
$\pi$ component is shown as a dashed line). The value 
$\Delta x$ is a distance between two $\sigma$ components,
which needs to be measured to determine the field strength.
By rotating a 
tip plate (panel e), the observer brings two oppositely polarized  
components in line with each other (panels f and g), 
and the tip angle is used to derive the field 
strength. 
}\label{fig:zef}
\end{figure*}

A daily observation of sunspot field strengths starts by the observer 
creating a full disk drawing of sunspots. Next, the observer centers each spot 
on the spectrograph slit and takes the measurement. 
Figure \ref{fig:zef} provides a graphical explanation of the measuring steps.
The spectrograph is 
equipped with a device that contains two strips of optical retarders 
(quarter-wavelength plates) placed above each other in the direction along the spectrograph slit. 
The two retarders have their primary axes at a 90-degree angle with respect to each other 
and are set to be at 45 degrees to the prime axis of the Nicol prism polarizer 
situated behind the 
retarders (Figure \ref{fig:zef}b). The detailed description of this setup can be found in 
\citet{Hale1913,Ellerman1919,Hale.Nicholson1938}. In the presence of a longitudinal magnetic field in 
the solar photosphere, a spectral line (sensitive to the presence of the magnetic field)
splits into two components, in which one component has a left-hand circular 
polarization and the other component is circularly polarized in the
right-hand sense. The polarization pattern depends on the polarity of the magnetic field. Passing through the quarter 
wavelength plate converts the circularly polarized light to linearly
polarized with two orthogonal orientations, which are  spatially 
separated by the Nicol prism. In the absence of a transverse component, the 
observer sees the two 
components of the Zeeman triplet above each other, spatially shifted 
relative to each other (Figure \ref{fig:zef}d). The above setup 
has no 
significant effect if the incoming light has linear polarization due to transverse Zeeman effect. 
For linear polarization, the central $\pi$ component of spectral 
line is visible in both spectra, as depicted in Figure 
\ref{fig:zef}c. 

The separation between two $\sigma$ components represents the total field strength.
The polarization pattern of the components (circular, linear, or a mix of both -- elliptical)
contains information about the orientation of the magnetic field vector relative to the line of sight.
Thus, the linear separation $\Delta x$ (mm) between the two circularly polarized $\sigma$ components can be used to 
measure the (total) field strength, $B$ (G) as follows:

\begin{equation}
B = \frac{s~ \Delta x \cdot 10^{13}}{9.34 \cdot g~\lambda^2},
\label{eq:zeeman}
\end{equation}
where $\lambda$ is the wavelength (in \AA), $g$ is 
Land{\`e} factor, and $s$ is the linear dispersion (\AA/mm; see Eq. \ref{eq:disp}).

To measure the separation $\Delta x$ between the two 
$\sigma$ components of the spectral line, the spectrograph 
is equipped with a transparent 
(glass) plane-parallel plate (the so-called tip plate), which 
can be tilted (Figure \ref{fig:zef}e). Light from the low
part of the spectrum passes through this tip plate, and tipping 
the plate allows the observer to align the two $\sigma$ 
components (Figure \ref{fig:zef}f, g). Equation 
\ref{eq:ppp} shows the relation between the rotation angle 
$\alpha$ of the tip
plate and the lateral separation $\Delta$x of spectra, i.e.,

\begin{equation}
\Delta x = t\cdot\sin\alpha~\Bigg(1 - \frac{\cos \alpha}{\sqrt{n^2 - \sin^2 \alpha}}~\Bigg),
\label{eq:ppp}
\end{equation}
where 
$t$ is the thickness of the plate and 
$n$ is its index of refraction.
The rotation angle $\alpha$ is calibrated to $\Delta$x, 
and, typically, the values are written in a look-up 
table for the observer to translate the tip angle into the corresponding field 
strength. The polarity of the magnetic field is determined from the 
direction of the lateral shift required to align the $\sigma$ 
component in the lower part of spectra with the upper part
(see Figure \ref{fig:zef}g). In the early period
of MWO observations, instead of look-up tables, the observer used
the tip angle directly as the measure of the field strength; i.e., the angle in degrees
corresponded to a field strength in units of 100 G. For example, 10$\degr$ would correspond to 1000 G, 25$\degr$ would be recorded as 2500 G, etc. 
Using a tip plate as the measuring device has two disadvantages. First, 
the dependence 
between the tip angle and the lateral 
shift is nonlinear (Equation \ref{eq:ppp}). 
The nonlinearity is relatively small for
small tip angles, but it increases rapidly for large angles.
Second, the range of measurements is restricted by the 
maximum tip angle of the plate (about 60$\degr$). This limits the maximum field strength that could be measured.

\begin{figure} 
\centerline{\includegraphics[width=1.\columnwidth,trim=0 0 74 100,clip=]{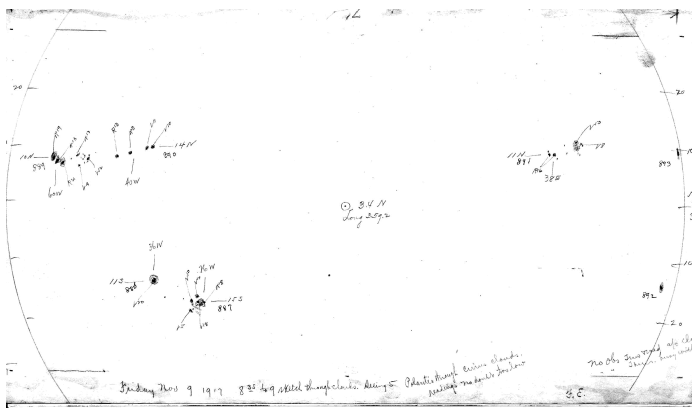}}
\caption{(a) Example of a typical sunspot drawing from the MWO dataset taken on 9 November 1917
between 8:35 and 9:00 (local time). 
Measured sunspot field strengths are shown in units of hundred 
G and their polarities designated as ``R'' and ``V'' (for red and violet, which corresponds to the positive and the negative polarities). Credit: Carnegie Observatories.
}\label{fig:drawing}
\end{figure}

Figure \ref{fig:drawing} provides an example of one of the earliest 
sunspot drawings. For each sunspot, the observer would draw the umbra 
and penumbra; the umbra is shown as the dark-shaded area and the penumbra is shown 
as a light-shaded area or outlined by a contour. In addition to 
the sunspots, their magnetic polarity and field strength, the drawings show the location of 
the solar disk center with its approximate latitude (solar B-angle) 
and the solar limbs. Other notations on the drawings include the date and time of 
observations, the atmospheric seeing conditions (smaller number 
corresponds to poorer seeing conditions), and the observer's initials. While the characterization 
of the seeing conditions is subjective, for a reference we note
that, based on a personal experience of one of the coauthors, the 
observations denoted as ``Seeing 3'' were taken on a partially cloudy day 
with the solar granulation clearly visible. The seeing scale is between 1 
and 5.

Some drawings may include other handwritten notes either related to the 
observations (e.g., indicating that some sunspots were measured on a 
different day owing to the weather condition), or even personal notes; for example, there is a note about the first observations, taken with the then world's largest 100 
inch (2.5 m) Hooker telescope, planet Mercury transit on 7 May 1924. Many drawings 
made during the first two months after the project started are not annotated with field strengths However, the same 
drawings published in \citet{Hale.Nicholson1938} show field 
strengths measured in these sunspots. We think that for the early 
period, the measured field strengths were not written on the 
drawings, but were perhaps recorded in some other way. 
Since our database is based on information from drawings, these cases
are currently not included.

The quality of the drawings shows a significant evolution over the period of the program. 
\citet{Hale.Nicholson1938}, in their reference to 1917--1924 period, 
indicated that the drawings should be 
considered as approximate, chiefly for the purpose of the identification of 
sunspots for which the measurements were taken. Indeed, the drawings from 
that early period appear to be less rich in detail, and may often not even show 
multiple umbrae inside sunspots. In contrast, the more recent drawings 
exhibit an extremely 
high degree of detail and, while they look like photographs, in some 
respect they are even better, as a skilled observer can draw the details 
based on moments of clearer seeing. The level of details in drawings 
made in other periods vary between these two extremes. Normally, it 
takes about 10-15 minutes for the  current observer to complete the 
drawing, but the notes written on drawings suggest that sometimes it 
took longer (e.g., because of clouds or rapid changes in weather). 
Perhaps as an indication of uncertainty in time of observations,
the observers sometimes recorded the time of observations rounded to the closest
15 minutes. This time quantization could be identified from the beginning of
observations, but it became universally adopted after about 1979.
Cases when the drawing had to be interrupted and completed at a 
later time are normally indicated with the appropriate time of observations
for each group of features. Nevertheless, the users of these drawings 
need to understand that the locations of sunspots on a drawing are not 
recorded exactly simultaneously, and the features may have slightly shifted from the 
location corresponding to the indicated time of drawing as a consequence of solar rotation.

The drawings contain a sign of a developed understanding of new solar phenomena. 
For example, the early measurements had uncovered a 
possible polarity reversal of the magnetic field of sunspots
situated near the solar limb. It was clearly studied 
in detail as the drawings from that time period would often exhibit a line dividing 
sunspots into two parts, with the field strength and the polarity measured 
in each part (e.g., 15 January\ 1918, sunspots near the east-north limb). 
Some drawings may show a 
set of lines across the sunspot with the measurements taken along each line across the sunspot (e.g., 22 July 1918). 
It appears that at some point, the origin of this phenomenon was finally understood
\citep{Hale.etal1919}, and drawings after that time no longer 
include such details. As we know now, this effect is due to a 
projection of a fan-like structure of magnetic field in a sunspot onto 
the line-of-sight direction. The database includes both measurements of the main 
polarity and the polarity appearing due to projection. 

As other example, drawing taken on 19 November 1921 
shows marks for measured magnetic field to the west of a large unipolar sunspot.
This was the result of a search for ``invisible'' sunspots of opposite polarity
next to unipolar spots \citep{Hale1922}. The database does not include these measurements outside sunspot or pores.

Initially, the observations were taken in spectral line \ion{Fe}{I} 
6173.343 \AA\ (Land{\'e} factor g=2.5). \citet{Hale.Nicholson1938} 
quoted the 
width of the spectrograph entrance slit as 76~$\mu$m, and the 
observations were taken in the second spectral order of the spectrograph. 
This slit width is close to the normal slit width, which is the minimum width that
maximizes the resolving power of the spectrograph without sacrificing the intensity 
of spectra. An examination 
of observer logbooks indicates that there was  no standard protocol  for recording such information as the spectrograph settings and, when 
changes were made, they could be recorded in various ways (or were 
not recorded at all). Thus, for example, \citet{Ulrich.etal1991} 
reported that for sunspot field strength measurements a slit of 356 
$\mu$m width is used. Recent examination of the slit width 
has revealed an even larger width of about 500~$\mu$m; there is no recording of when this 
change was made. This indicates that the slit 
adjustments were not well documented.

Beginning October 1961, the observations were switched to the spectral 
line \ion{Fe}{I} 5250.217 \AA\ with the Land{\'e} factor of g= 3.0. The 
measurements were made in the fifth spectral order. This switch required 
replacing the previous 4 mm\footnote{According to \citet{Ellerman1919} in 1919, the thickness
of tip plate was one eight of an inch, or about 3 mm.} thick tip plate (with the index of refraction of $n$ = 1.521 for crown glass) by a thicker tip plate of 7 mm (with the index of refraction of $n$ = 1.461 for fused silica). 
This replacement introduced the limitation on the maximum strength of magnetic field that could be measured to approximately 3000 G. 

There were nine spectral gratings over the lifetime of the project. 
Properties of these gratings are summarized in Table \ref{tab:disp}; 
additional details can be found in \citet{Livingston.etal2006}. Spectral 
orders are not well recorded in observer logs. Thus, we estimated spectral orders
on the basis of scaling used to convert tip angles to gauss and the linear 
dispersion computed by us using parameters of gratings. The spectrograph 
grating equation for the Littrow configuration could be written as 
\begin{equation}
m\lambda = 2d~ \sin\beta,
\label{eq:grating}
\end{equation}
where $\beta$ is angle of diffraction, $m$ is the spectral (diffraction) order, $d$ is the spacing between the grooves (i.e., $d^{-1}$ represents a more familiar measure of number of grooves per millimeter), and $\lambda$ is the wavelength. The linear dispersion ($s$) is 
\begin{equation}
s = \frac{\partial \lambda}{\partial x} = \frac{d~ \cos\beta}{mL},
\label{eq:disp}
\end{equation}
where $L$~=~22.9 m is the effective focal length of the spectrograph. Using 
Equations \ref{eq:grating} and \ref{eq:disp} we computed 
the linear dispersion for all spectral gratings employed in 
the course of MWO sunspot field strength project.
Table \ref{tab:disp} 
shows the measured magnetic field strengths (100~G, 1000~G, 2000~G, 
3000~G, and 4000~G) and the ``true'' strengths that were   
computed using Equations \ref{eq:zeeman}-\ref{eq:ppp}. The true field strengths are rounded
to the nearest 10 G. For measurements taken after 1961, the true field strengths that require 
tip angles in the excess of 60$\degr$ are not shown in Table \ref{tab:disp}. For 
measurements taken after 1994, the lookup table was used.  
The lookup table does not include values for 
tip angles smaller than 22$\degr$ and thus the 
entry for 100 G in Table \ref{tab:disp} is not shown.

While these calculations should be considered as approximations, they clearly demonstrate that 
the measuring method adopted at MWO slightly 
overestimates the weaker field strengths, and it could 
significantly underestimate the strong magnetic fields.
As noted by \citet{Livingston.etal2006}, 
before 1961 the linear dispersion in the observing 
wavelength range changed very little, and the 
resulting differences between the measured and true fields (computed from the Equations \ref{eq:zeeman}-
\ref{eq:ppp}) are within 100 G (with the exception 
of strongest fields $\ga$ 3500 G; see also, Table \ref{tab:disp}).
After October 1961, when the observations switched to the spectral line \ion{Fe}{I} 525.0217 nm (and the 7 mm tip plate),
the measuring method was to divide the tip angle 
by two and round to it to the nearest degree \citep{Livingston.etal2006}. 
Analyzing Equation \ref{eq:ppp}) suggests that 
this method works reasonably well for gratings no. 
6--8 and for field strengths $\la$ 2000 G.
During this period, there is a strong nonlinearity for magnetic fields in the range  2500-3000 G (measured fields). 
For observations taken after 1994 (grating no. 9),
a 
lookup table was used to convert the tip angles to the 
field strengths. The lookup table, which is currently (2018) used for the observations is shown in Table \ref{tab:obs}. For 
comparison, we show the range of true values of magnetic fields corresponding to the measured fields and rounded to the nearest 100 G.
\citet{Livingston.etal2006} noted that the lookup table 
of measured field values overestimates the weaker fields ($\la$ 2200 G) 
and underestimates stronger fields ($\ga$2700 G). For data 
taken during this period, \citet{Livingston.etal2006} proposed 
corrections for the measured field strengths. 
For 1917-1961 period, our calculations for true field strength agree 
with \citet{Livingston.etal2006}. In reference to post-1994 
measurements using the lookup table, \citet{Livingston.etal2006} wrote that for the maximum tip angle of 60$\degr$, the lookup table indicates 2600 G while the true field 
strength is about 2960 G. Our calculations for this case, rounded to the 
nearest 10 G also show 2960 G.
However, for measurements during 1961--1994, \citet{Livingston.etal2006} 
listed only a single correction, while there were three different spectral 
gratings with slightly different linear dispersion. 
Thus, for example, the correction Table II in
\citet{Livingston.etal2006} showed that the published field of 3000 G
corresponds to true field of 3800 G. Our calculations return 4050 G 
(grating no. 6), 3870 G (grating no. 7), and 3480 G (grating no. 8).
Another reference point mentioned in \citet{Livingston.etal2006} is a tip
angle of 32$\degr$ during 1961--1994 period, which was said to correspond 
to an actual field strength of 1680 G. Our calculations return 1620 G 
(grating no. 6), 1550 G (grating no. 7), and 1400 G (grating no. 8).
With respect to early data, 
according to \citet{Hale.Nicholson1938}, only measurements above 
1000 G are reliable; those below 1000 G are merely an approximation.

\section{Digitization}\label{sec:digit}

The drawings were scanned using a flat-bed scanner and are available online  
\footnote{\url{ftp://howard.astro.ucla.edu/pub/obs/drawings/}}. Scanning was done by the 
observatory personnel between their 
main duties. Unfortunately, different scanner 
settings were used in different periods and thus 
we had to set the scaling to the actual size of the scanned images.

The conversion of information from the drawings was done 
in several simple steps. First, the operator 
identified the location of the solar limbs and the 
center of the solar disk. The code analyzes the 
intensity in vicinity of these points to improve 
the identification of the solar limbs. Then the limbs and the 
solar disk center are fitted by a circle in image coordinates. This 
establishes the coordinate system relative 
to the image and, as the next step, each sunspot is manually marked 
by the operator. The date and time of 
observations are entered manually. 
The solar ephemerids are calculated 
and  used to transform the image coordinates of 
sunspots to the heliographic coordinates, i.e., solar latitude and the central meridian 
distance (CMD). For each marked sunspot, the operator manually enters the 
polarity and field strength as recorded on the drawing; no 
correction, for example, for the nonlinearity of the tip plate is made. If desired, 
the user is able to make her or his own corrections to the field strength using 
the information provided in this paper (Equations 
\ref{eq:grating}--\ref{eq:disp}, Tables 
\ref{tab:disp}--\ref{tab:obs}) or as proposed by \citet{Livingston.etal2006}.

After the operator identifies the center of each sunspot, the code 
analyzes the intensity of nearby pixels attempting to identify a 
continuous dark-shaded area of sunspot umbra. The pixels identified 
as belonging to the umbra are used to calculate an 
estimated area of umbra. Caution is necessary 
when using the areas of sunspot umbra
derived from the drawings as these data 
may not be optimal for quantitative studies.

The drawings are made on a standard drawing size 
paper sheets, which normally include only a 
portion of solar disk between approximately 
$\pm$35$\degr$ in solar latitude. Sunspots at higher latitudes, 
which fall outside the drawing range are plotted on the same drawing 
as insets. Usually, the insets are identified by a small box 
drawn around its boundaries, and the heliographic coordinates 
(latitude and longitude) of the center of the inset are identified. 
In a few cases, the insets may correspond to the same active 
region observed on several days (even though the rest of the drawing 
may correspond to a single day). To digitize insets, the operator would 
first mark the center of the inset and enter its 
heliographic coordinates as shown on the drawing. This 
establishes a secondary coordinate system, which is used to calculate 
the correct heliographic coordinates of sunspots included in the inset.

Drawings taken prior to about mid-1960s show the time of observations 
in local time. All drawings made after 27 February 1969 show 
Universal time (UT). During the transition, the time could be 
recorded either in local time or UT depending on personal preferences
of the observers. Unfortunately, the time designation (UT, AM, or PM) 
was missing from some drawings.
For consistency, the database preserves the original time recorded 
on the drawing. The conversion from local to UT time is done 
separately. 
Although there were several early periods when daylight 
savings time (DST) was used in California, the MWO may 
have continued using Pacific standard time (PST) even when the whole 
country adjusted the clocks ahead by one hour after the Congress passed an 
act ``to promote the national security and defense by establishing daylight
saving time''.
Thus, we interpret all
local time shown on the drawings as PST. This seems to be confirmed by rare instances, when the drawings list PST as the time reference (e.g., 5-18 April 1918 or 16-18 July 1943). 
For a reference,
Table \ref{tab:dst} 
lists historical time periods when the DST was in 
effect in the state of California. As a test of the validity of our determination of time
designation (AM, PM, and UT), we conducted a statistical analysis of 
the distribution of time of observations to verify that the observations 
occurred when the Sun was above the horizon. The database 
includes both time taken from the drawings and UT time.
To convert the time of observations shown on original drawings to UT, future users 
simply need to add 8 hours independent of time of year. Unfortunately, there are examples in which the conversion was done assuming DST in California \citep[e.g., see Figure 2 in][]{Lundstedt.etal2015}.

This digitization project included all drawings taken from 1917 till 
2016. No 
drawings after 2016 were digitized. The total number of drawings in our dataset is 36714 and the total number of features 
(sunspots and pores) identified on these drawings is 470544.

The data extracted from the drawings were placed  in a fully 
searchable database based on an open-source relational 
database management system MySQL. Each drawing is represented by 
a set of the following parameters: date and time of observations (both local and UT), atmospheric seeing, 
observer's initials, observer's notes, 
radius of solar disk in pixels, the image x,y 
coordinates of the center of the solar disk in pixels, 
P angle (orientation of the solar north relative to terrestrial 
north direction, degrees), B angle (the latitude of the center of 
solar disk, degrees), Carrington longitude of the solar central 
meridian (degrees), number of sunspots on a drawing, and the extracted (or expected to be extracted)
data for each 
sunspot as listed in Table \ref{tab:mysql}. The database is 
expandable to include other parameters (e.g., MWO or NOAA group number)
or a similar type of data from other instruments.
Figure \ref{fig:alldat} shows the field strengths of all measured 
sunspots.

\section{Known issues with measurements}\label{sec:issues}

\begin{figure} 
\centerline{\includegraphics[width=1.\columnwidth,trim=80 210 90 210,clip=]{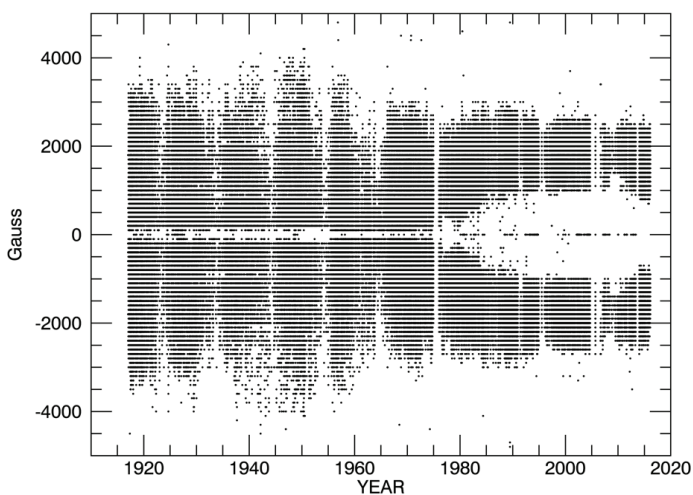}}
\caption{Field strengths (dots) measured in all sunspots included in our digitization project.
}\label{fig:alldat}
\end{figure}

The final database has several deficiencies of different 
origins. Thus, for example, we can see an upper limit in the field strengths 
measured after 1961 (following the switch to spectral 
line \ion{Fe}{I} 525.0217 nm; see Section \ref{sec:method}). 
A slightly lower upper limit can be identified in post-1994 data.
According to the lookup Table \ref{tab:obs}, there should be no 
measurements of field strengths above 2600 G after 1994. However, the drawings 
taken during this period  show field strengths of 2700 G. 
Based on communication with the current observer (Steve 
Padilla), the largest sunspots may show field strengths, which 
appear to be outside of the measuring range (60$\degr$ of tip 
plate). For these sunspots, the field strength is recorded as 2700 G.
A few data points in post-1994 data, which show even larger field strengths in 
Figure \ref{fig:alldat}, are due to digitization errors.

Another prominent 
feature is the absence of weak field measurements in the
later part of the dataset. Prior to mid-1970, it was not 
uncommon for the observer to measure sunspot field strengths 
as low as 100 G. Between 1975 and 1985, the data 
show a clear trend with fewer 
weak field measurements then in previous years. 
According to Larry Webster, the head observer at that time, 
``this trend is associated with two big changes 
related to replacing the Babcock grating [no. 7, Table 
\ref{tab:disp}] by Bausch and Laumb grating [no. 8, Table 
\ref{tab:disp}] following the pit flood and the restoration 
of the original 8-inch Littrow lens. These two changes produced 
a significant improvement in 
the quality of the spectrum at the final eyepiece and, as the result, 
[he noticed that] all spots that could be resolved on days of good 
seeing showed fields of 1500 G or more. It took a while for all the 
observers to have enough good seeing experience that they agreed 
[with him], so there was not a jump [but a gradual change] in the minimum 
field immediately following the two above upgrades.''
We note, however, that the trend of fewer weak field sunspot
measurements starts in about 1976 -- well before replacing the
grating no. 8 (1982, Table \ref{tab:disp}). 
Interestingly enough, the observations of sunspot field strengths 
taken at the CrAO \citep[][and references therein]{Pevtsov.etal2011} also show a 
development of a similar gap in weak fields. Unlike the MWO observations, 
however, the gap in CrAO data already appeared in about 
1965-66 and, by 
early 1970s, the measurements of fields less than 1000 G became 
extremely rare. By contrast, sunspot 
field strength measurements from the Main (Pulkovo) Astronomical 
Observatory \citep[][and references therein]{Pevtsov.etal2011} 
exhibited a gap in field 
strengths since the beginning of that dataset (1957). This gap 
(or minimum) in weak field strength measurements, which is 
subjective, may demonstrate the impact of a somewhat 
new knowledge (e.g., expectations 
from theoretical studies) on measurements, when the observer 
consciously tries making 
the observations to fit his or her understanding of a physical 
phenomenon (i.e., no sunspots should have a field strength 
smaller than 1000--1500 G).

Figure \ref{fig:alldat} also shows the presence of a slight observational (quantization) bias in the measured field 
strengths. This bias can be identified by following one of the 
horizontal streaks in Figure \ref{fig:alldat}. Thus, for example, the
density of points at 2000 G is higher than the density of measurements in two neighboring 
streaks (1900 and 2100 G). This bias further reveals itself in a  
histogram of field strengths (not shown) as small local peaks 
in the number of measurements at (both positive and 
negative) 800, 1000, 1500, 1800, 2000, and 2200 G.  
In bins neighboring these values (i.e., 900, 1100, 1400, 
1600, 1700, 1900, and 2100 G), the number of 
measurements is lower. The origin of this bias is unknown, 
but we might speculate that it could be due to a
subjective preference, when the observer rounded up 
the measured tip angle to the ``nearest'' integer angle.
This bias is present in both 1917-1960 and 1961-2016 
subsets of data, but it is stronger in the former.

\citet{Hale.Nicholson1938} indicated  
that only measurements above 1000 G are reliable.
\citet{Livingston.etal2006} also noted that the
measurements of magnetic field strengths weaker than 1000 G
are unreliable because in the weaker fields the Zeeman splitting for \ion{Fe}{I} 6173.343 
\AA\ becomes comparable to the Doppler width of the spectral
line. 
On the other hand, the same article \citep[][]{Livingston.etal2006}
countered this by a statement that a well-trained observer can
consistently measure fields weaker than 1000 G, and by providing an 
example, where multiple measurements by the same MWO observer 
agree within 100 G. 

While we tend to agree that the measurements of B$\la$ 500-1000 G should be treated with a caution, the polarity measurements appear to be very robust. In a few selected cases, we 
checked the polarity of features
with weak magnetic field during their disk passage, but we did not find 
a random reversal of polarity, which should have occurred if those field strengths were unreliable (i.e., below a 
detection threshold). 
Nevertheless, in our analysis of drawings, we came across with
a very small number 
of erroneous polarity measurements when the polarity of a  
same sunspot changed as it crossed the solar disk.

\section{Errors and uncertainties from digitization}\label{sec:errors}

When the original drawings were scanned by the observer at MWO, the 
scanning area was deliberately reduced by cutting off what the 
scanner operator considered as unessential information. Thus, for 
example, the scanned area often stops
exactly at the limbs (and sometimes, limbs near the equator were
excluded). In some cases, the scanned images may exclude one of the solar 
limbs altogether; this is probably because no sunspots were present in that part
of the Sun. These omissions make it harder to establish a proper 
orientation of drawings with respect to their E-W orientation. The omissions 
may also 
lead to larger uncertainties in fitting the solar limb. The scanned images from  
1996 till early 2000 do not include the date and time of observations. 
The latter omission hinders the proper calculations of the Carrington longitude.
The original drawings contain information on the 
sequential number of each drawing, the number of groups, and (sometimes) the number of sunspots. 
There is also information on the orientation of the drawing (angle) relative to 
the direction of the solar equator. Unfortunately, this information was 
excluded from the scanned area.
One of the coauthors (AAP-2) 
visited the archives of Carnegie Observatories in summer 2018 and photographed the 
drawings of 1996-2000 that miss this metadata. These photographs are used
to update the MySQL database with the correct metadata, although this update has not
been completed yet.
Our visit to the archive of Carnegie Observatories uncovered that images of 
scanned drawings available online may be different from the drawings preserved 
in the archive. Some of these discrepancies could be explained by the processing
of drawings at MWO, which involved several steps as follows:
\begin{enumerate}
    \item 
The spots are drawn using the projected solar image as a guide.
    \item
The drift angle is measured by offsetting the image in an EW direction. The date and time were then entered to calculate the p angle.
    \item 
The polarities and field strength are measured with the tipping plate.
    \item 
The group type is identified from the distribution of the polarities and 
 configuration of the spot parts.
    \item 
The center of gravity as judged by the observer is identified and noted by 
vertical and horizontal tic marks.
    \item 
The positions of the spot groups are determined from the Stonyhurst charts 
and noted on the drawing.
    \item 
The observation time is entered on the drawing along 
with the observers ID, observing conditions, and any other comments.
The intent was to record the mid-point of the drawing process but because the time had been entered to get the drift angle that time was occasionally used.
    \item
After 1996 the drawing was scanned and posted to the web page.
    \item
At the end of the month the drawings for the month were reviewed by one of the senior observers, obvious errors were corrected, either erased or crossed out. Typically these errors are of a typographic nature.
    \item
The final parameters
of each spot group were entered into the monthly sunspot report and the actual drawings were stored in the archive of the Carnegie Observatories in Pasadena.
The final parameters are online at the MWO web page and they are also
available from Solar Geophysical Data (SGD)
\footnote{\url{ftp://ftp.ngdc.noaa.gov/STP/SOLAR_DATA/SUNSPOT_REGIONS/Mt_Wilson/}}.
These reports are available for the years between 1962 and 2004.
\end{enumerate}

Between 1996 and early 2000 the scanner in use for the promptly posted
drawings was too small, resulting in the loss of critical metadata on these 
posted images.  This metadata (time and date of observations) is available
from the SGD reports and the entries in that publication are definitive.  
Copies of spots that were outside the scanned
area were digitally relocated to the page area and labeled as an inset.  The 
relocated tic marks and Stoneyhurst positions
are from the actual locations in such cases. Examples of such insets can found 
in 15 May 1999 and 29 August 1999 scanned drawings.
Owing to the review process, we caution that there may be differences
between the posted images and the definitive publication.


Since the digitization of the drawings was done using manual 
identification,  there could be misidentifications and some features 
could be completely missed. In addition, there are errors that were 
made when the information from the drawings was entered 
manually during the digitization process. This includes errors in 
the time of observations, sunspot polarity, and field strength.
We evaluate the level of errors using entries by two students.
The students processed over 28,000 images spanning 1917 through 
2017. Their task was to enter the metadata (date and time of 
observations, coordinates of numbered groups identified on drawings, 
seeing, and comments). In an effort to expedite the work the students were assigned either odd or 
even years. As a benchmark of the quality of their work, both students 
processed the same drawings for years 1920, 1940, 1960, 1980, and 
2000.  Examining these overlapping years allows us to 
understand the level of possible human errors.    
The first error arises in the number of images processed. There were 
1493 images spanning these five overlapping years.  One student processed all images 
while the other only processed 1481, thus missing 12 images or about 
0.8\% of images in this sample. Of the 1481 images that they both processed there 
were 51 differences in hour, minute, time designation (AM, PM, or UT), or seeing on 35 
images, which is an error rate of 0.86\% limited to approximately of 2.3\% 
drawings. These errors included those that were caused by poor handwriting, 
multiple observation times,  and seeing values shown on the drawings, swapping hours and minutes 
and errors with no identifiable reason.  
The analysis of errors indicates that the quality of manual digitization 
improved as students gain the experience working with the data. Noticeably,
fewer mistakes were made in relation to multiple and missing entries for time and seeing 
conditions. This could also be attributed to better policies being adopted with regard 
to the information being recorded on the drawings. Handwriting varies 
wildly between the observers and definitely affects the amount of 
mistakes. Knowing the observers, it 
might be possible to identify those observers whose drawings
have a higher risk for mistakes during the process. A bigger 
problem is the amount of unforced errors, for example, errors with no noticeable 
cause, especially during the later overlapping years (1980 and 2000).  
 
From the analysis of ``number of errors by year'', the big worry is 
the number of unforced errors (unknown cause), which include omitted or incorrect times (hours, minutes, seconds, 
AM/PM/UT designation) and seeing quality. However, overall the data quality improved 
with considerably fewer errors recorded from 1960 onward.  
From the analysis of ``type of errors'', most errors  
occurred only once except for errors with no obvious (unknown) cause and poor 
handwriting.
There is a 
marked difference in that one of the students made many more of these unforced 
errors.
Breaking down the errors by year, we can see that one student was 
fairly consistent with unforced errors, committing at least one 
error in every year except for 1960.  In contrast, the other student 
committed all of his unforced errors on 1980 data.  

Assuming that quality of the data and the error rate remained consistent 
across the entire dataset we would expect to see a total of 884 drawings 
with at least one error, of which 249 are due to an error 
introduced by the human mistake, such as swapping hours and minutes or 
other error (unknown cause).
Owing to early oversight, the length 
of the comment field was limited to 100 characters, which in some cases had truncated a long comment.


The criteria for the digitization 
program had evolved as the project developed. In the early period of the project, sunspots without the field strength measurement (marked only by a letter corresponding to their 
polarity) were not included in the digitized database. Later, these sunspots were
added, but to accommodate the input format, the operator would have to assign either 
$+$49 or $-$49 (the maximum field strength allowed by the Kislovodsk digitization program). The database contains about 6.3\% of entries with $\pm$4900 G. 
While majority of these entries correspond to sunspots with only polarity (no field 
strength) measurement, there could be a few sunspots whose field 
strength was that value. Future updates will correct this uncertainty.

Digitization made by the NSO team saved the image 
coordinates 
of all features (the disk center, limb marks, and sunspot locations), 
but the digitization
made by the Kislovodsk team only stored image coordinates for the 
disk center. For other features, only the heliographic coordinates 
were preserved. The image coordinates could be useful in case
the heliographic coordinates need to be updated.
The image coordinates could be computed using the heliographic 
coordinates of sunspots, but the conversion depends on scaling of the 
scanned images. Unfortunately, after the digitization was completed 
it was discovered that some original images stored on MWO server had 
been modified by adding comments to the digital images. This changed 
the size of images and inadvertently broke the scaling between the 
image and heliographic coordinates. Future users of this
dataset may notice such cases when the recomputed image coordinates 
of marked features (e.g., sunspots) may be offset relative to 
the features on a drawing. 

In summer 2018, one of the coauthors (AAP-2) examined the original drawings held in the 
archive of the Carnegie Observatories and discovered 
several drawings (from 1996--2000 period), which were not scanned. Thus, these 
data are absent from the digitized dataset. It is currently unclear 
how many drawings were not scanned, but we do not expect it to be 
a large number.

These deficiencies will be addressed in the later editions of the 
dataset, subject to available resources.
To keep track of future modifications to the original digitized dataset,
we introduce the version number. The results of the initial 
digitization already published in some early papers
\citep[e.g.,][]{Tlatova.etal2015,Tlatova.etal2018} are designated as
version  MWO\_SPOTS\_20171101, 
where the numbers represent the year, month, and 
day of last changes to the dataset. The current version of the
dataset described in this paper is MWO\_SPOTS\_20180801.


Uncertainties in the heliographic coordinates of sunspots may arise from an incorrect 
determination of image coordinates of the solar features, position of the solar disk center, 
and radius of the solar disk on drawings. As one example of an  
observer error,  the first drawing (4 January 1917; see online archive of sunspot drawings at 
\url{ftp://howard.astro.ucla.edu/pub/obs/drawings}), the CMD of the leading sunspot of the Mount Wilson group 516 
is denoted as {\rm 9W}, while in fact, the correct CMD should be {\rm 19W}.

One source of  uncertainty can be associated with the thickness of lines drawn by the 
observer. Limited examination of drawings for different years indicates
that a typical thickness of lines varies between
5 and 7 pixels (in scanned image scale). For comparison, a typical radius of the 
solar image is about 1250 pixels. Figure \ref{fig:rsun} provides an example of variations in radius of the solar disk (R$_{sun}$) as measured from the drawings. As 
expected, R$_{sun}$ shows an  annual variations of about $\pm$1.6\%. 
Amplitude of R$_{sun}$ in pixels depends on scaling that was 
selected during the image scanning. We note a deviation
between the fitted sine curve and the measured R$_{sun}$ in mid-1982, which 
has been traced to a use of slightly different resolution for a flatbed 
scanner. A significant deviation in early 1979 (vertical line reaching very low
values of R$_{sun}$) is also due to a significantly different scanner scale
selected by the observer. Smaller spikes 
in R$_{sun}$ are random uncertainties in the radius of the circle fitted to the solar disk.
The deviations of  the measured R$_{sun}$ from the fitted sine curve are 
normally distributed and have a zero mean and standard deviation of 
$\sigma$ = 4.5 pixels.
The nonuniformity in selected scaling for scanning 
the drawings can be clearly seen in Figure \ref{fig:rsun_all}.
Uncertainties in solar latitude and 
CMD of sunspots arising from effect of the above uncertainty in R$_{sun}$ 
on image-to-heliographic coordinate 
transformation vary with heliographic distance from the solar disk 
center. For heliocentric distances $\rho \le$ 70$\degr$ the uncertainty is smaller than 
0.10 degrees. For $\rho \approx$ 75$\degr$ the uncertainty reaches 0.15 degrees. 
At $\rho \approx$ 85$\degr$ the uncertainty is about 0.5 degrees, and near 
the limb ($\rho \approx$ 87$\degr$), it increases to more than one degree. 
The above uncertainties correspond to one standard 
deviation derived from 1000 realizations of coordinate transformations 
using six-pixel uncertainty in the position of the solar limb.

The drawings do not include northern or southern parts of 
the solar disk at high latitudes and, thus when determining the radius of the solar disk, only eastern and western 
limbs were used. This might introduce a minor uncertainty due to 
a distortion of the (assumed round) shape of the solar disk by atmospheric refraction. 
However, for a reasonable zenith angles, the effect is small 
(less than 1 arcsecond), when compared with the average angular diameter of solar disk of 
about 1800 arcseconds \citep[see Figure 4 in ][]{Corbard.etal2013}. Even a larger 
uncertainty in solar limb could result from 
the atmospheric seeing. However, a typical daytime atmospheric seeing at 
MWO is about 2-3 arcseconds, which approximately corresponds to 3-4 pixels on
a typical drawing. Thus, the overall effect from both 
atmospheric refraction and atmospheric seeing is comparable to the 
uncertainty associated with the line thickness of drawings.

Uncertainties in the heliographic coordinates of sunspots 
may also arise from a slight displacement of sunspots during the 
time period when the observer records sunspots on a drawing. 
During this time, the spots may change 
their positions because of solar rotation, which could add as much as 
0.15 degrees of uncertainty to their observed longitudes (assuming 
that it takes 15 minutes to complete the drawing). The 3 arcsecond 
atmospheric seeing could result in $\approx$ 0.18 degree 
uncertainty in measured position of a sunspot.
Based on the above arguments, we conclude that with the exception of
errors in identifying sunspots, 
the uncertainties in the heliographic coordinates do not exceed 0.5 degrees.

\section{Comparison with other datasets}\label{sec:compare}

To compare the sunspot field measurements from different instruments, 
we selected 100 sunspots 
observed over the 20-year period between 1994 and 2014 both at MWO and CrAO. Sunspots that were observed on the same day by both observatories 
were selected for comparison. To ensure that the same magnetic features were compared, we excluded
sunspots with multiple umbrae or a complex magnetic pattern. The sunspots were also 
selected to represent a broad range of field strengths.

The program of measuring sunspot field strengths at Crimea 
started in mid-1950s and (with some brief interruptions) has continued until the present \citep{Pevtsov.etal2011}. The measurements are made 
visually by several staff observers, which is the same practice as in past MWO observations; 
more recent observations at MWO are taken by a single observer.
 Additional information
about the quality of CrAO sunspot field strength measurements can be found in  
\citet{Lozitska.etal2015}. 
In a search for possible systematic trends, we divided the dataset in two equal halves: 
50 sunspots observed during 
1994--2003 and 50 sunspots in 2004--2014. Figure \ref{fig:mso_crao} shows 
a scatter plot of measurements from one observatory versus the other. The 
measurements from the two observatories correlate 
reasonably well, Spearman rank correlation coefficient is 0.64 with
(probability of no correlation) $p$ = 6 $\times$ 10$^{-13}$. 
The local time difference between the two observatories is ten hours, which may partially explain
the scatter between the measurements of the same sunspots in two observatories.
The least-squares linear  
fit (assuming 100 G uncertainties in both MWO and CrAO measurements)
yields B$_{MWO}$ = (549.00$\pm$82.23) + (0.73$\pm$0.04) B$_{CrAO}$
\citep[see][for a discussion of the fitting approach]{Nagovitsyn.etal2016}. Slope of fitted 
line is steeper for the early period (1994--2003),   
B$_{MWO}$ = (436.45$\pm$134.84) + (0.79$\pm$0.06) B$_{CrAO}$ 
than for the (later 2004-2014), 
B$_{MWO}$ = (646.88$\pm$102.30) + (0.67$\pm$0.05) B$_{CrAO}$.

For another comparison, we used the total field strengths derived by fitting the 
Stokes I profiles observed by the Vector Stokes Magnetograph (VSM) on 
Synoptic Optical Long-term Investigation of the Sun (SOLIS) platform 
\citep[e.g.,][]{Balasubramaniam.Pevtsov2011}. We carried out the fitting  via the SOLIS 
Zeemanfit code \citep{Hughes.etal2013}, which employs a three-component 
fit to the non-polarized (Stokes I) profile of \ion{Fe}{I} 6302.5 \AA\ spectral line. 
In sunspots with a strong magnetic field, the Zeeman splitting becomes wide enough 
for the triplet nature of the line to be clearly visible in non-polarized light 
(for example, see Figure \ref{fig:zef}a). Therefore, a three-line fit to the spectra 
provides a fairly robust measure of the total magnetic field 
strength. The field strengths using Zeemanfit are computed routinely beginning 
in late-2013. Figure \ref{fig:mso_solis} shows a scatter plot of MWO sunspot 
field strength measurements versus SOLIS Zeemanfit for 50 sunspots 
observed between October 2013 and December 2014. For this comparison, sunspots were selected 
on the basis of closeness in observing time between the two instruments and to ensure a broad 
range of field strengths in a sample. As it is difficult to identify the exact location of 
a flux element for which the MWO measurements were taken, for this comparison we avoided sunspots with complex
polarity patterns to minimize spatial misidentification of measured flux elements.
Two data sets show 
a good relationship with Spearman rank correlation $r$=0.85 and $p$= 5.6 $\times$ 10$^{-15}$.  The least-squares linear 
fit returns B$_{MWO}$ = (1054.86$\pm$89.81)+(0.45$\pm$0.03) B$_{SOLIS}$.
For this fit, we assumed the uncertainty in B$_{MWO}$ measurements to be 100 G.
For SOLIS data, the uncertainties are returned by the Zeemanfit code, 
which 
for this subset of data were between 47~G and 251~G.
In this comparison, we took the MWO field strengths as 
they are shown on the drawings without any correction for a nonlinearity in the
measured fields.

In absence of evolution, the magnetic field strength in a stable sunspot should 
exhibit very limited center-to-limb variation. As a test, we selected 
several well-developed sunspots and followed the variations in measured field strength
during their disk passage. As a general rule, the field strength exhibits clear 
maximum when the sunspot is near the central meridian. In all studied cases, the decrease 
in observed field strength as function of the heliographic distance from solar
disk center is less steep as compared with the cosine function. On average, a difference 
between sunspot field strength near the central meridian and the heliocentric distance of 
50$\degr$ is about 300 G, which is in agreement with the center-to-limb 
variation in the height formation of the photospheric spectral lines and known vertical 
gradient of the magnetic field in the photosphere \citep[1~G/km,]{Borrero.Ichimoto2011}. 

\section{Conclusions}\label{sec:conclude}

The results reported in this article concern the digitization of historical 
sunspot field (strength and polarity) measurements made at MWO. 
The systematic observations began in 1917 and have continued until the present with a 
few brief interruptions. These manual measurements represent the earliest 
observations of the magnetic fields on the Sun and, while they may have many 
shortcomings when compared with the modern polarimetric observations, these data 
provide a critical link with past magnetic activity of the Sun. Given the long-term duration of this time series, we think that its  
continuation would be extremely important. Based on comparison with 
CrAO and SOLIS/Zeemanfit measurements, we argue that the time series could be 
successfully continued using the data from these two observatories in 
case the sunspot operations at MWO should end.

The digitized dataset may still contain some errors related to the  orientation of
drawings or to an incorrect identification of  solar limbs and/or sunspot features. 
There could also be a few errors related to the time of observations. These errors will 
be corrected in future releases of the dataset, once the team becomes aware 
of the specific errors. 

%

\begin{figure} 
\centerline{\includegraphics[width=1.\columnwidth,clip=]{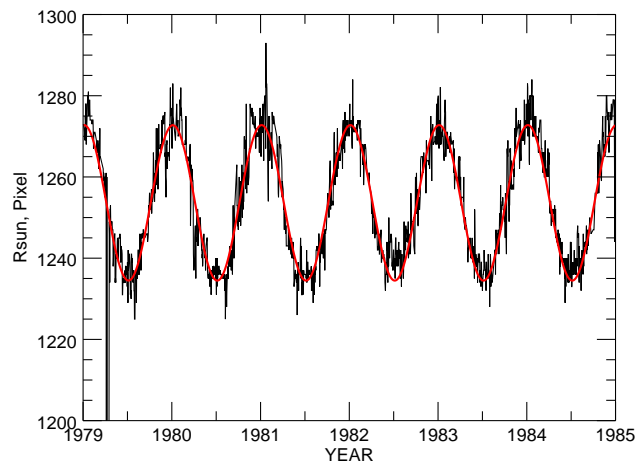}}
\caption{Radius of the solar disk in 1979--1982 as measured from drawings (black solid
line). Red line shows the best least-squares fit by a sine function.}
\label{fig:rsun}
\end{figure}

\begin{figure} 
\centerline{\includegraphics[width=1.\columnwidth,trim=80 210 90 210,clip=]{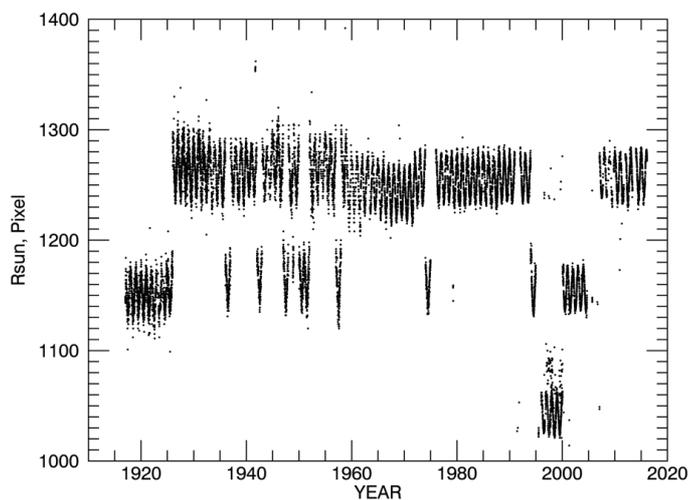}}
\caption{Radius of the solar disk for all dataset as measured from drawings.}
\label{fig:rsun_all}
\end{figure}

\begin{figure} 
\centerline{\includegraphics[width=1.\columnwidth,clip=]
{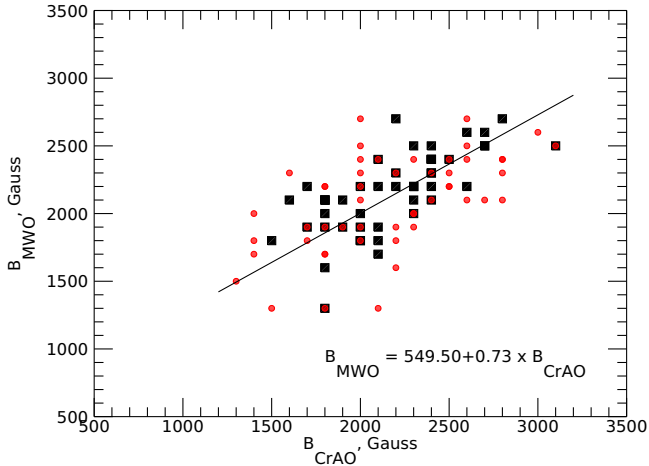}}
\caption{Comparison of same-day measurements of field strengths of 100 sunspots observed 
at MWO and CrAO. Observations 
from 1994--2003 are shown by black squares, and data for 2004--2014 are shown 
as filled red circles. The solid line indicates the least-squares fit by the first degree polynomial with 
errors in both MWO and CrAO data. The coefficients of best-fit line for all data are shown in 
low-right side of the Figure.
}\label{fig:mso_crao}
\end{figure}

\begin{figure} 
\centerline{\includegraphics[width=1.\columnwidth,clip=]{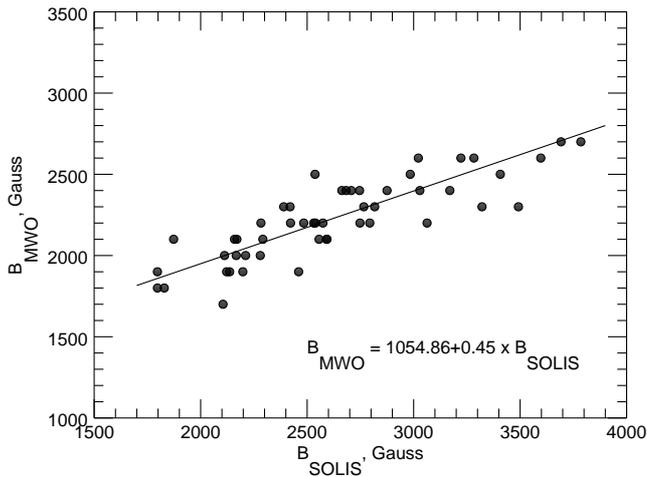}}
\caption{Comparison of field strengths of 50 sunspots observed at MWO and by 
VSM/SOLIS (Zeemanfit code). Solid line is a least-squares 
fit to the data by a first degree polynomial. Parameters of fitted 
polynomial are shown in the lower part of the plot.}
\label{fig:mso_solis}
\end{figure}

\begin{figure}
\centerline{\includegraphics[width=1.\columnwidth,clip=]{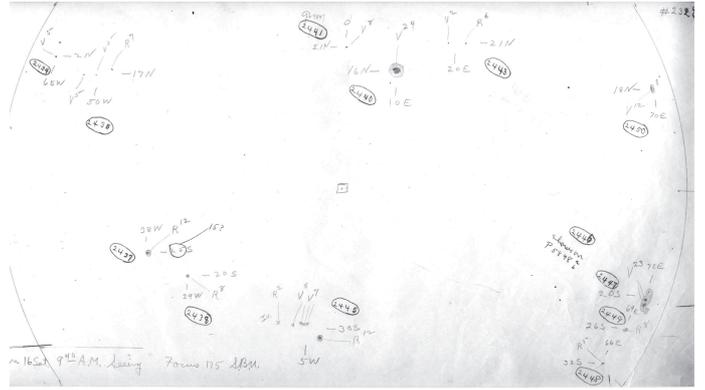}}
\caption{Example of drawing taken on 16 January 1926. Sunspot outlines lack fine details. Scanned
image excludes portion of the drawing with year and month of observations and part of solar limb.}
\label{fig:appndx1}
\end{figure}

\begin{figure}
\centerline{\includegraphics[width=1.\columnwidth,clip=]{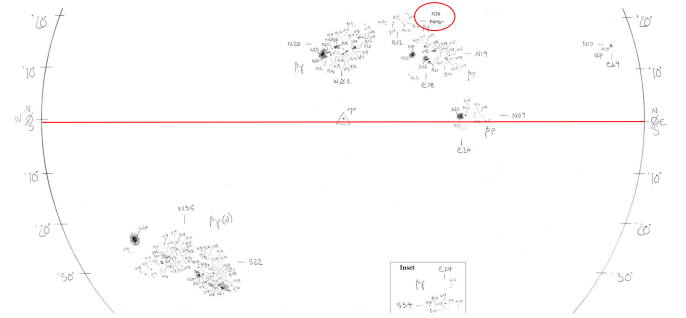}}
\caption{Example of drawing taken on 29 August 1999 showing high degree of 
detail. The scanned image excludes portion of the drawing with information about 
date and time of observations. The red line shows the horizontal direction in image 
orientation. Comparing the location of end points, it is clear that there is a slight misalignment between the horizontal direction and east-west
orientation. Portion of the image enclosed by box is an inset, which shows 
portion of solar image not in its correct location on solar disk. The red oval 
indicates an example of an error in the heliographic coordinates of an active region. In this particular case, the later correction of latitude of this region is in error; compare the number with the approximate latitudes indicated on the solar limbs.}
\label{fig:appndx2}
\end{figure}

\begin{figure}
\centerline{\includegraphics[width=1.\columnwidth,clip=]{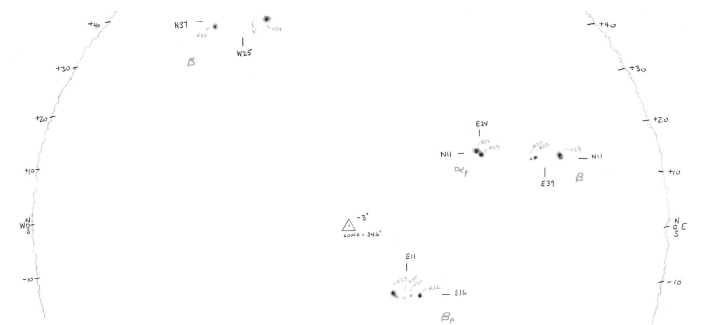}}
\caption{Example of drawing taken during the period of a poor seeing. The location of solar limb is shown approximately, using "wavy" lines.}
\label{fig:appndx3}
\end{figure}

\begin{figure}
\centerline{\includegraphics[width=1.\columnwidth,clip=]{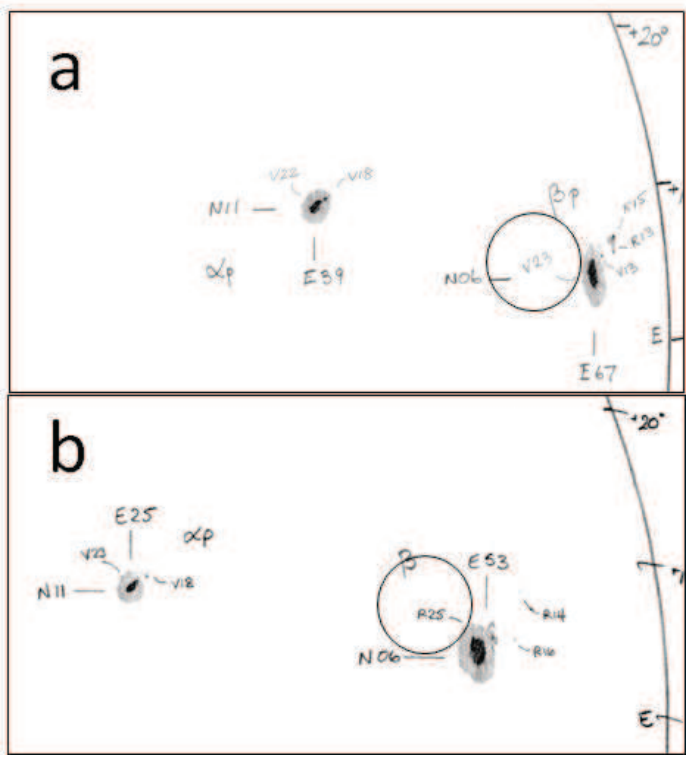}}
\caption{Example of incorrect measurement of sunspot polarity. Polarity of a sunspot observed on 13 November 2013 (panel a) was determined as "V" (negative), but on all following days (panel b, 14 November 2013), the measured polarity was "R" (positive). The circles indicate the polarity measurement for this sunspot.}
\label{fig:appndx4}
\end{figure}

\begin{figure}
\centerline{\includegraphics[width=1.\columnwidth,clip=]{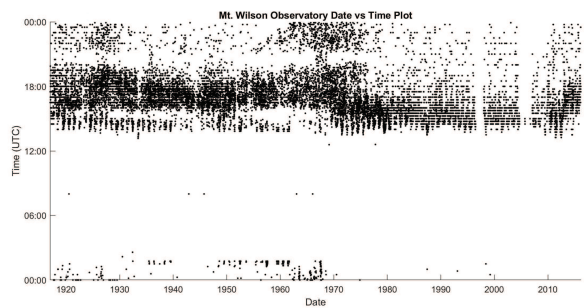}}
\caption{Universal time of observations for all drawings in the dataset. The vertical gap in 1996-1997 corresponds to period when the time of observations was not included in the scanned
images. These data will be updated in later versions of the 
dataset. A second vertical gap corresponds to 2004--2007, when the 
sunspot drawings program was shut down. Annual variations
in the earliest time of observation are due to change in the sunrise time during the year. Two horizontal gaps in seeing in the time of observations correspond to a shading of the coelostat mirror tracking the Sun by the support pillar and secondary mirror (which reflects light to the main telescope mirror) in early morning hours (about 15:00 UT) and the break for the lunch hour at the MWO (about 20:00 UT). The lunch break was a formal event at MWO until about 1984,
when the operations of solar telescopes were transferred 
from Carnegie observatories to UCLA. Early morning horizontal 
gap (15:00 UT) seems to disappear after 1970, which suggests 
that at that time,
the morning observations may have been 
taken with the first flat mirror 
moved to the east position, and thus having a strongly oblique angle between the two mirrors.
}
\label{fig:appndx5}
\end{figure}

%
\begin{table*}
\caption{Parameters of spectrograph gratings used during the lifetime 
of project}\label{tab:disp}
\centering
\begin{tabular}{crlclrrccccc}
No.&Start date&grooves&$\lambda$&$m$&$s^{-1}$&Scale\tablefootmark{a}&\multicolumn{5}{c}{Measured and true\tablefootmark{b} field strengths (G)}\\
&&mm$^{-1}$&\AA&&mm/\AA&100~G&100&1000&2000&3000&4000\\
\hline\hline
1&1917&602&6173&2&2.97&1$\degr$&90&920&1900&3010&4320\\
2&1930&602&6173&2&2.97&1$\degr$&90&920&1900&3010&4320\\
3&May 1949&400&6173&3&2.96&1$\degr$&90&920&1900&3020&4340\\
4&29 Sep 1950&400&6173&3&2.96&1$\degr$&90&920&1900&3020&4340\\
5&19 May 1955&600&6173&2&2.96&1$\degr$&90&920&1900&3020&4340\\
6&22 Aug 1960&600&6173&2&2.96&1$\degr$&90&920&1900&3020&4340\\
same\tablefootmark{c}&Oct 1961&600&5250&5&11.15&0.5$\degr$&90&940&2180&4050&...\tablefootmark{d}\\
7&24 Dec 1962&610&5250&5&11.66&0.5$\degr$&90&900&2080&3870&...\tablefootmark{d}\\
8&17 May 1982&632&5250&5&12.96&0.5$\degr$&80&810&1870&3480&...\tablefootmark{d}\\
9&21 Nov 1994&367.5&5250&9&15.27&Table \ref{tab:obs}&...\tablefootmark{e}&760-800&1880-1940&...\tablefootmark{e}&...\tablefootmark{e}\\
\hline
\end{tabular}
\par\noindent
\tablefoottext{a}{Angle of rotation of tip plate per 100 G.}
\tablefoottext{b}{Computed using Equations \ref{eq:zeeman}-\ref{eq:ppp}.}
\tablefoottext{c}{Grating No. 6 was used from 22 August 1960 through 23 December 1962. In October 1961, the wavelength
for sunspot observations was changed to \ion{Fe}{I} 5250.217 \AA.}
\tablefoottext{d}{Out of measurement range for tip plate.}
\tablefoottext{e}{Out of range of lookup Table \ref{tab:obs}.}
\end{table*}
\begin{table}
\caption{Conversion table used by the observer for measurements taken after 1994.}\label{tab:obs}
\centering
\begin{tabular}{ccc}
Tip angle&Measured&True\\
deg.&field, G&field, G\\
\hline\hline
22-23&1000&800\\
24-26&1100&800-900\\
27-28&1200&1000\\
29-30&1300&1100\\
31-32&1400&1100-1200\\
33-35&1500&1200-1300\\
36-37&1600&1400\\
38-39&1700&1500\\
40-42&1800&1600-1700\\
43-44&1900&1800\\
45-46&2000&1900\\
47-48&2100&2000-2100\\
49-51&2200&2100-2300\\
52-53&2300&2300-2400\\
54-55&2400&2500-2600\\
56-57&2500&2700\\
58-60&2600&2800-3000\\
\hline
\end{tabular}
\end{table}
%
\begin{table*}
\caption{Historical periods when the DST was in effect in Los Angeles, California
(years after MWO adopted the UT timing practice are not included)}\label{tab:dst}
\centering
\begin{tabular}{llll}
Years&Start date&End date&Note\\
\hline\hline
1918-1919&31 Mar&27 Oct&Introduced by ``Standard Time Act of March 19, 1918''\\ 
&&&abolished by 66th U.S. Congress in "CHAP. 51.\\
&&&An Act for the repeal of the daylight-saving law"\\
1942-1945&9 Feb 1942&30 Sep 1945&year-round DST by Public Law 403 "Daylight Savings Time"\\
1949-1966&last Sunday in Apr&last Sunday in Sep&DAYLIGHT SAVING TIME California Proposition 12 (1949)\\
1967-1972&last Sunday in Apr&last Sunday in Oct&U.S. Uniform Time Act of 1966\\
\hline
\end{tabular}
\end{table*}

%
\begin{table*}
\caption{Example of entries reserved for parameters extracted from the drawings}\label{tab:mysql}
\centering
\begin{tabular}{rl}
No.&Parameter description\\
\hline\hline
1&latitude of sunspot, degrees\\
2&CMD of sunspot, or ``longitude'' relative to central meridian, negative/positive \\
&east/west of central meridian, degrees.\\
3&longitude (Carrington longitude of sunspot, degrees\\
4&heliocentric distance, r/R, where r – distance from disc center, R – radius of solar disk\\
5&Reserved for: area of sunspot, millionths of solar hemisphere\\
6&Reserved for: area of sunspot, pixels\\
7&Reserved for: size of sunspot in latitudinal direction, degrees (could be negative as it is computed as difference in latitudes).\\
8&Reserved for: size of sunspot in longitudinal direction, degrees\\
9&Reserved for: average intensity\\
10&Reserved for: contrast of sunspot relative to surrounding photosphere\\
11&Reserved for: minimum intensity in sunspot\\
12&Reserved for: maximum intensity\\
13&Reserved for: tilt angle relative to equator, degrees\\
14&field strength (in units of 100 G)\\
15&Reserved for: length of outer boundary of sunspot, pixels\\
\hline
\end{tabular}
\end{table*}
%
\appendix   
\section{Example of uncertainties arising from quality of hand drawings}

Uncertainties in heliographic coordinate transformation may arise 
from several sources including uncertainties in fitting solar limb, image position
of sunspots, orientation of scanned image, and uncertainty in time when 
the position of a specific sunspot was recorded.

When the modern observations of sunspot drawings are taken, the observer
orients the drawing paper in the direction parallel to solar equator (see,
Section \ref{sec:errors}). This is done manually by eye under variable atmospheric seeing conditions, and thus inherently, the image orientation must have some uncertainty. The exact amplitude of this uncertainty 
depends on the observer's level of training and the observing conditions, 
but it is reasonable to assume that it should not exceed about one degree (larger image inclinations are usually detectable by eye). This error 
mainly affects the heliographic latitude, and it increases with the 
distance from the disk center. For each individual observation, the error 
is systematic, which  means\ that the latitudes of sunspots located near one solar limb 
are slightly larger than the true latitude, while the latitudes of sunspots located near 
opposite limb are slightly smaller. 
It might be possible 
to estimate (and correct) this error in orientation of daily images, 
by comparing the latitudes of sunspots in sequential daily observations
assuming that sunspot latitude does not change as it crosses solar visible disk. Alternatively, we can estimate the error in heliographic
coordinates for each sunspot location on the disk using, for example, 
Monte Carlo simulations. Based on the above considerations, the 
sunspots situated at the solar limb should have an uncertainty in their 
latitude of about one degree or less. For sunspots near the disk center the 
uncertainty in latitude is much smaller.

The image orientation could have been further altered during the drawings scanning (i.e., placing the drawing slightly tilted relative to scanner bed). This error in image orientation could be readily detected 
(and corrected) if the scanned image includes both solar limbs with markings 
corresponding to location of solar equator (see, Figure 
\ref{fig:appndx2}). Unfortunately, number of scanned images may exclude 
either one of the limbs or its near equatorial part (Figure \ref{fig:appndx1}).
In the latter case, the errors in heliographic coordinates can be evaluated on the 
basis of Monte Carlo simulations assuming some mean value of tilt of images based on other scanned images.

In the worse case scenario, two errors (errors in orientation due to observing conditions
and due to image scanning) could combine, thus doubling the 
uncertainties in latitude of sunspots close to solar limbs.

The above procedure of orienting the drawing paper in respect to solar equator was not always followed, and thus there are several cases in 
early part of the dataset when the 
drawings do not have a proper orientation (e.g., drawing orientation was not corrected for P angle). While we tried determining the proper image 
orientation for such cases, there could be some instances that were not corrected. 
Those should become evident when the derived parameters of sunspots are used to 
investigate their evolution during solar disk passage with following visual inspection 
of individual images. Once detected, such cases should be used to update the database.

Uncertainties due to unknown time of observations have a larger effect on the heliographic longitudes.  Usually, the time recorded on the drawing corresponds to 
the beginning of observations although in some cases the drawing may show a time range 
(beginning and end time of drawing). Based on more recent observations, typically the 
time to complete a drawing may vary between 15 and 30 minutes although it may take 
longer depending on the level of observer's training, atmospheric conditions, and sunspot activity; for example, we compare the number of features 
marked by the observer on drawings shown in Figures \ref{fig:appndx1} and 
\ref{fig:appndx2}. Based on the initials of the observers, in early part of the dataset,
it was not uncommon to have an observer taking observations only for one or two years 
and using substitute observers (nonsolar members of MWO scientific and supporting 
staff, and even observatory visitors). Thus, it is very possible that some drawings 
taken in early period of the dataset could take longer than 30 minutes.
Figure \ref{fig:appndx5} shows UT time of drawings for 1917--2016 period. After 
about 1979, the time shows quantization, when the observers started rounding up the time to the 
nearest 15 minutes. Similar quantization could also be seeing in some early periods 
although it is not as pronounced as in post-1979 period. Time quantization
may create a visual impression of fewer observations taken after 1979. This impression is
incorrect. The number of observations per year is about the same during these two periods. 

During 1967-68, there are larger uncertainties in 
the time of observations. It appears that more
drawings did not show ``PM'' or ``AM'' 
designation, and some observers may have been 
experimenting with using UT time. The time of 
observations for this period may require  
additional examination.

Within the period each drawing was taken, it is 
unknown which sunspot was drawn at what time. This uncertainty in time of drawing for 
each feature introduces uncertainty in their heliographic 
longitude. Owing to solar 
rotation, in half an hour the sunspot location could change by 
about 0.3$\degr$ depending on sunspot latitude. Figure \ref{fig:appndx5} 
shows that for several
drawings the time of observations converted to UT may be 
incorrect. In rare cases, the drawings were taken late in the 
day when the UT time may exceed 24:00. This could  result in an 
error in the day of observations; the day of observations is 
recorded as a calendar day, not UT day.
This will be verified and corrected 
in future releases of the dataset. The uncertainty in time should be 
considered when the sunspots from this dataset are used for determination of solar rotation 
and sunspot proper motions.

At the beginning of the dataset, the drawings were considered as approximate sketches 
and did not contain fine details (Figure \ref{fig:appndx1}). Images taken in later 
periods show extremely high level of detail (see, Figure \ref{fig:appndx2}). This 
difference in degree of detailization may affect the uncertainties in image 
location of sunspots as well as their (sunspot) areas 
with higher uncertainties for more approximate images. The exact 
amplitude of these uncertanties may be hard to estimate, but as a low limit, we can
assume that they should not be smaller than the thickness of a pencil used by the 
observer. Based on our analysis of width of solar limb on selected drawings, it is about 
5--7 pixels in scanned images (see, Figure \ref{fig:rsun}).
This value can be 
used in combination with the size of sunspots and their location on the disk as well as 
the uncertainty in the location of fitted solar limb and radius of solar disk on a 
drawing to derive a more precise estimate of uncertanties in heliographic coordinates of each feature.

Some drawings may contain insets -- a portion of solar image drawn not in its right 
location (see Figure \ref{fig:appndx2}). Typically, insets were created by the observer to 
accommodate high-latitude groups falling outside of drawing paper. Some insets were created 
during the scanning process, as a solution to a limited size of a scanner bed.
The center of sunspot group in the inset is marked by the 
observer (or scanner operator), and to convert image coordinates to the heliographic coordinates, we 
have to use these heliographic coordinates first to convert 
image coordinate of the insert to 
its proper location on the drawing, and then convert new image coordinates to proper 
heliographic coordinates. Such a double coordinate transformation provides only approximate correction for foreshortening effects, and thus this transformation may introduce uncertainties
in relative heliographic coordinates of sunspots shown in the inset. Moreover, the 
heliographic coordinates shown on insets are approximate (rounded to a closest degree), 
which further increases uncertainties of heliographic coordinates of features shown 
as insets. The amplitude of the effect depends on heliographic position of the inset,
and thus needs to be evaluated separately for each sunspot in the inset. We expect, 
however, that the resulting errors will be smaller than the errors resulting from 
uncertainties in radius of solar disk (limb fitting) and image orientation.

The location of solar limb is marked by the observer as part of making a 
drawing. Fitting circle to the hand-drawn solar limbs results in additional 
uncertainty in solar radius. However, we find such uncertainty is comparable 
to a thickness of pencil lines. When the observations are taken in poor seeing
conditions, the observer may indicate the location solar limb by a "wavy" 
lines as shown in Figure \ref{fig:appndx3}. For such days, the user should expect to 
see larger errors in heliographic coordinates, but this needs to be evaluated using 
Monte Carlo simulations with larger uncertainties both in radius of solar image and the 
location of sunspots on images. For this modeling, the amplitude of wavy pattern could 
be used as a measure of uncertainties in solar disk radius and image position of sunspots.

Figures \ref{fig:appndx1}-\ref{fig:appndx2} show examples when some 
metadata (e.g., date and time of observations) could be excluded during the 
scanning of a drawing. Once identified, such cases should be used to update the database.

In some rare occasions, the measured polarity of sunspots could be in error. The 
exact number of such erroneous measurements is hard to estimate, as they  
require a detailed examination of sequential drawings. We identified only a very 
small number of such errors.

Overall, our assessment is that the uncertainties in radius of solar disk, 
image orientation, and time of observation are the largest source of uncertainties in 
the heliographic coordinates. In general, the uncertainty in coordinates of sunspots is within 1--2\degr.

The magnetic field measurements show several systematic effects, which may be impossible 
to fully account for. The minimum uncertainty of each measurement is 100 G, but for some 
measurements, the errors could be much larger.

%
\begin{acknowledgements}
The dataset is available under DOI: 10.25668/bkt9-4d24.
Sunspot drawings used in this work with permission of the Mt. Wilson 150-foot solar tower project, UCLA.
Authors are grateful to the countless observers, whose dedicated work 
resulted in the creation of a dataset spanning over more than 100 years.
Thanks are also due to the current volunteer observer, Steve Padilla, 
whose dedication keeps this invaluable dataset alive. 
We are thankful to the staff of the Carnegie Observatories and especially Dr. Cynthia Hunt for 
providing us (AAP-2) access to the archive containing the MWO drawings and 
for their 
dedication and support during our visit to Pasadena, California.
Thanks are due to two high school students: Teemu Nikula 
and Hilla Saukko for their help with verifying the metadata from 
the drawings.
This paper has benefited from discussions with Dr. Jack Harvey.
We acknowledge the financial support by the Academy of 
Finland to the 
ReSoLVE Centre of Excellence (project no. 307411). 
Also acknowledged 
is the funding from the Russian Foundation for Basic 
Research (RFBR, 
project 18-02-00098) and the Russian Science 
Foundation (RSF, project 
15-12-20001). US contribution 
to this project was partially supported by NASA NNX15AE95G grant.
The authors are members of the international team on Reconstructing Solar 
and Heliospheric Magnetic Field Evolution Over the Past Century 
supported by the International Space Science Institute (ISSI), 
Bern, Switzerland. The National Solar 
Observatory (NSO) is operated by the Association of Universities for 
Research in Astronomy (AURA), Inc., under cooperative agreement with 
the National Science Foundation.

\end{acknowledgements}

\bibliography{pevtsov_bibliography}
\end{document}